\documentclass[reprint,prl,aps,amsmath,amssymb,superscriptaddress]{revtex4-2}

\usepackage{times}	
\usepackage{amsmath,amssymb, amsthm}   
\usepackage{bm}
\usepackage{graphicx}
\usepackage{units}
\usepackage{hyperref}
\hypersetup{
  colorlinks=true,
  citecolor=blue,
  urlcolor=blue,
  linkcolor=red,
  breaklinks=true
}

\begin{document}


\title{Moiré-driven equilibrium of perturbations in moiré systems}

\author{Federico Escudero}
\email{federico.escudero@imdea.org}
\affiliation{IMDEA Nanoscience, Faraday 9, 28049 Madrid, Spain}

\author{Zhen Zhan}
\email{zhen.zhan@imdea.org}
\affiliation{IMDEA Nanoscience, Faraday 9, 28049 Madrid, Spain}

\author{Pierre A. Pantale\'on}
\email{pierre.pantaleon@imdea.org}
\affiliation{IMDEA Nanoscience, Faraday 9, 28049 Madrid, Spain}

\author{Francisco Guinea}
\affiliation{IMDEA Nanoscience, Faraday 9, 28049 Madrid, Spain}
\affiliation{Donostia International Physics Center, Paseo Manuel de Lardiz\'{a}bal 4, 20018 San Sebastián, Spain}

\begin{abstract}
Perturbations in moiré materials, such as due to substrates or strain, are common in many experiments and can significantly modify the electronic properties of the system. Here, we show that perturbations in twisted bilayer graphene tend to be transferred between the coupled Dirac cones, eventually reaching an equilibrium near the magic angle. We connect our results to experiments and show that this equilibrium behavior remains robust even when the moiré potential itself is perturbed. Our findings extend the notion of the magic angle to a more general regime governed by moiré-driven equilibrium. 
\end{abstract}

\maketitle

\textit{Introduction}— 
The remarkable properties of multilayer moiré materials~\cite{andrei2021marvels}, such as unconventional superconductivity~\cite{cao2018unconventional, yankowitz2019tuning, oh2021evidence} and strongly correlated phases~\cite{cao2018correlated, kerelsky2019maximized, xie2021fractional}, are intrinsically tied to the moiré coupling between the stacked layers~\cite{Pong2005review, brihuega2012unraveling}. In the particular case of twisted bilayer graphene (TBG)~\cite{andrei2020graphene}, the moiré coupling results in the formation of flat bands around the so-called \textit{magic angle}~\cite{lopes2007graphene, suarez2010flat, trambly2010localization, shallcross2010electronic, bistritzer2011moire, san2012non}. The strong enhancement of electron correlations in these flat bands leads to a rich set of correlated phenomena observed in numerous experiments~\cite{lu2019superconductors, sharpe2019emergent, saito2020independent, serlin2020intrinsic, zondiner2020cascade, stepanov2020untying, choi2021correlation, cao2021nematicity, stepanov2021competing}.

It is therefore important to understand the nature of flat bands, and previous studies have identified many of their properties, such as topology~\cite{hejazi2019multiple, song2019all, song2022magic, navarro2025topological, paul2025emergent}, origin~\cite{Tarnopolsky2019Origin, wang2024role, escudero2024diagrammatic}, and the recurrence of magic angles~\cite{Navarro2023Magic,Tarnopolsky2019Origin, becker2022mathematics}. However, the behavior of perturbations at the magic angle has remained largely unexplored. Perturbations in moiré heterostructures are ubiquitous in experiments and can significantly alter the electronic properties of the system~\cite{kazmierczak2021strain, choi2021correlation, de2022imaging}. For example, substrates such as hexagonal boron nitride (hBN) can lead to gapped Dirac cones~\cite{xue2011scanning, decker2011local, yankowitz2012emergence, zhang2019twisted, cea2020band, lin2020symmetry, shi2021moire, long2022atomistic, long2023electronic}, while relaxation- or strain-induced fields can shift the Dirac cones in both momentum and energy~\cite{van2015relaxation, naumis2017electronic, carr2018relaxation, guinea2019continuum, lucignano2019crucial, huder2018electronic, bi2019designing, escudero2024designing}. Interestingly, recent works have noted that the electronic properties near the magic angle appear to become insensitive to interlayer differences~\cite{long2022atomistic, yu2024twist, carrasco2025twistraintronics}.

In this work, we show that any perturbation in TBG is transferred to both layers at strong moiré couplings, eventually reaching an equilibrium state that we term \emph{moiré-driven equilibrium}. Using first-order perturbation theory, we show that layer perturbations are renormalized by the moiré coupling, in close analogy with the renormalization of the Fermi velocity in pristine TBG. This renormalization mixes the perturbations in the two Dirac cones, such that around the magic angle there exists a critical point at which they can become equal, regardless of their initial (decoupled) magnitudes. Our findings are supported by both first-order perturbation theory and extensive numerical calculations. By connecting our results to experiments, we further show that the moiré equilibrium tendency remains robust even when the moiré coupling itself is perturbed, thereby providing a natural explanation for why the individual properties of each layer become effectively masked, or difficult to disentangle, near the magic angle.

\textit{Model}— 
We consider TBG with a small rotation angle $\theta$. 
For a symmetric twist $\pm\theta/2$ in each layer, the reciprocal moiré vectors are given by $\mathbf{G}_{i}=\left[R\left(-\theta/2\right)-R\left(\theta/2\right)\right]\mathbf{b}_{i}$ ($i=1,2$), where $R\left(\theta\right)$ is the rotation matrix and $\mathbf{b}_{i}$ are the reciprocal lattice vectors of the honeycomb lattice~\cite{moon2013optical}. The $\xi$-valley ($\xi=\pm1$) Dirac points in the $\ell=t,b$ (top, bottom) layer read $\mathbf{K}_{\ell,\xi}=R\left(\pm\theta/2\right)\mathbf{K}_{\xi}$, where $\mathbf{K}_{\xi}=-\xi\left(2\mathbf{b}_{1}+\mathbf{b}_{2}\right)/3$ are the unrotated Dirac points.

\begin{figure}[t]
    \includegraphics[width=1\linewidth]{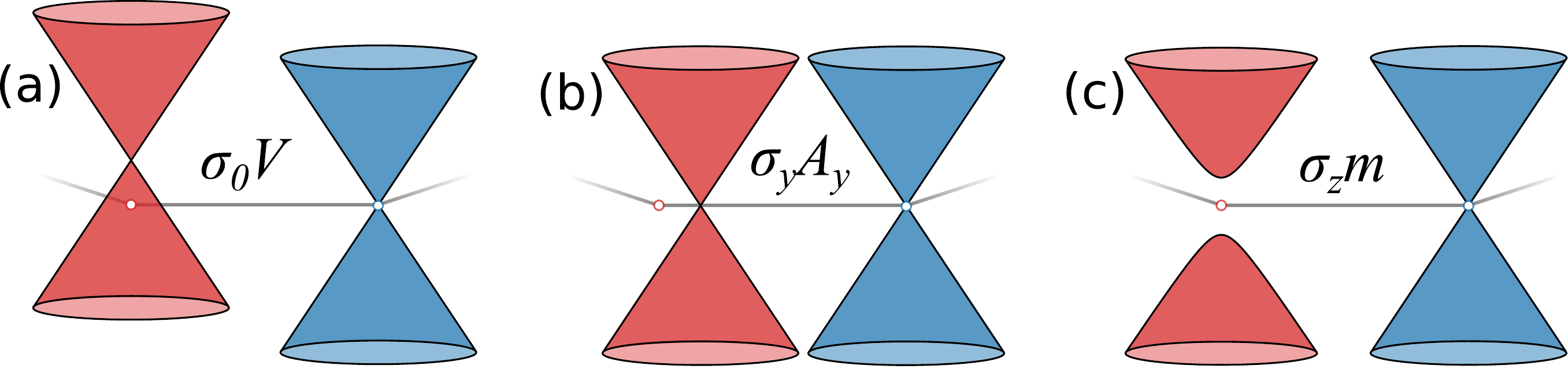}
    \caption{Schematic perturbation effect in TBG, for: (a) Scalar potential, (b) gauge potential and (c) mass potential. In all cases, in red and blue are shown the two Dirac cones at the $K_b$ and $K_t$ points of the moiré Brillouin zone (gray hexagon), coming from the twisted bottom and top layer, respectively. The perturbations shown only act on the bottom layer, moving the Dirac cones in (a) energy or (b) momentum, or (c) creating a bandgap. When coupled by the moiré potential, the Dirac cones hybridize and the perturbation is transferred between the layers.}\label{fig:perturbations}
\end{figure}

To capture the low-energy properties of TBG, we use a continuum model~\cite{lopes2007graphene, bistritzer2011moire, lopes2012continuum, moon2013optical, koshino2015interlayer, koshino2018maximally}. As intervalley scattering is negligible at low energies, we focus on the valley $\xi=+$ and drop the valley index for simplicity. The continuum model Hamiltonian reads 
\begin{equation}
H=\left(\begin{array}{cc}
H_{t} & U^{\dagger}\\
U & H_{b}
\end{array}\right),
\end{equation}
where $H_{\ell}=-\hbar v\boldsymbol{\sigma}\cdot\left(\mathbf{k}-\mathbf{K}_{\ell}\right)$ are the Dirac Hamiltonian relative to the twisted Dirac points and $U$ is the moiré potential. For simplicity, we preserve the particle-hole symmetry by neglecting the rotation of the Pauli matrices~\cite{bistritzer2011moire, moon2013optical, koshino2018maximally}, which does not modify our conclusions. The leading-order Fourier expansion of the moiré potential is given by~\cite{koshino2015interlayer,bistritzer2011moire} 
\begin{equation}
U=U_{1}+U_{2}e^{i\mathbf{G}_{1}\cdot\mathbf{r}}+U_{3}e^{i\left(\mathbf{G}_{1}+\mathbf{G}_{2}\right)\cdot\mathbf{r}},\label{eq:moire_pot}
\end{equation}
with
\begin{equation}
U_{j}=\left(\begin{array}{cc}
u & u'e^{-i\phi_{j}}\\
u'e^{i\phi_{j}} & u
\end{array}\right),
\end{equation}
where $\phi_{j}=\left(j-1\right)2\pi/3$, while $u$ and $u'$ are the effective AA and AB/BA hopping energies, respectively. 

We now consider a perturbation $P_{\ell}$ in each layer, so that $H_{\ell}\rightarrow H_{\ell}+P_{\ell}$. 
Examples of perturbations include perpendicular electric fields, substrate-induced potentials, and strain-induced modifications. A generic perturbation that incorporates these effects is
\begin{equation}
P_{\ell}=\sigma_0V_{\ell}+\boldsymbol{\sigma}\cdot\mathbf{A}_{\ell}+\sigma_{z}m_{\ell},\label{eq:Pl_particular}
\end{equation}
where $\sigma$ are the Pauli matrices and $\mathbf{A}_{\ell}=\left(A_{\ell x},A_{\ell y}\right)$. In the decoupled layers, the scalar potential $V_{\ell}$ shifts the Dirac points in energy~\cite{lopes2007graphene}, the in-plane parameters $A_{x},A_{y}$ shift the Dirac points in momentum space~\cite{Suzuura2002Phonons}, and the mass $m_{\ell}$ opens a gap at the Dirac point~\cite{Jung2015Origin}. A schematic illustration of these effects is shown in Fig.~\ref{fig:perturbations}.

\textit{First-order perturbation theory}— 
Insights on the perturbation effect can be obtained by truncating the continuum model Hamiltonian at the first shell, and performing first-order perturbation theory around the two Dirac points~\cite{bistritzer2011moire}. For small perturbations such that $\left|P_{\ell}\right| \ll \hbar vk_{\theta}$, where $k_{\theta}=\left|\mathbf{K}_{t}-\mathbf{K}_{b}\right|$ is the length of the moiré Brillouin zone, we can consider as a perturbation both the $\left(\mathbf{k}-\mathbf{K}_{\ell}\right)\neq0$ terms and the perturbations $P_{\ell}$. The perturbed Hamiltonian around both Dirac points then reads \cite{SM}
\begin{equation}
H_{\ell}^{\star}=-\hbar v^{\star}\boldsymbol{\sigma}\cdot\left(\mathbf{k}-\mathbf{K}_{\ell}\right)+P_{\ell}^{\star},
\end{equation}
where $v^{\star}=v\left(1-3\alpha^{2}\right)/\left[1+3\alpha^{2}\left(1+\lambda^{2}\right)\right]$ is the renormalized Fermi velocity~\cite{bistritzer2011moire}, with $\alpha=u'/\hbar vk_{\theta}$ and $\lambda=u/u'$ being the moiré coupling constant and the ratio between the AA and AB/BA hopping energies. The renormalized perturbation is given by
\begin{equation}
P_{\ell}^{\star}=\frac{P_{\ell}+\sum_{j=1}^{3}U_{j}h_{j}^{-1}P_{\ell'}h_{j}^{-1}U_{j}}{1+3\alpha^{2}\left(1+\lambda^{2}\right)},\label{eq:Pl_renorm}
\end{equation}
where $\ell'\neq\ell$ refers to the other layer and $h_{j}=\hbar v\boldsymbol{\sigma}\cdot\mathbf{q}_{j}$, where $\mathbf{q}_{j}$ are the three transfer vectors that determine the moiré Brillouin zone. The moiré potential renormalizes the perturbation to $P_{\ell}^{\star}$ by mixing the initial perturbations $P_{t}$ and $P_{b}$.

Equation~\eqref{eq:Pl_renorm} is valid for \emph{any }perturbation $P_{\ell}$. For the particular perturbation given by Eq.~\eqref{eq:Pl_particular}, the energies become
\begin{equation}
E_{\ell}=V_{\ell}^{\star}\pm\sqrt{m_{\ell}^{\star2}+\left|\hbar v^{\star}\left(\mathbf{k}-\mathbf{K}_{\ell}\right)-\mathbf{A}_{\ell}^{\star}\right|^{2}},\label{eq:El_pert}
\end{equation}
where
\begin{align}
V_{\ell}^{\star} & =\frac{V_{\ell}+V_{\ell'}3\alpha^{2}\left(1+\lambda^{2}\right)}{1+3\alpha^{2}\left(1+\lambda^{2}\right)},\label{eq:scalar_eq}\\
m_{\ell}^{\star} & =\frac{m_{\ell}+m_{\ell'}3\alpha^{2}\left(1-\lambda^{2}\right)}{1+3\alpha^{2}\left(1+\lambda^{2}\right)},\label{eq:mass_eq}\\
\mathbf{A}_{\ell}^{\star} & =\frac{\mathbf{A}_{\ell}-\mathbf{A}_{\ell'}3\alpha^{2}}{1+3\alpha^{2}\left(1+\lambda^{2}\right)}.
\end{align}
In the absence of perturbation, Eq. \eqref{eq:El_pert} reduces to the well-known Dirac dispersion $E_{\ell}=\hbar v^{\star}\left|\mathbf{k}-\mathbf{K}_{\ell}\right|$ with a renormalized Fermi velocity $v^{\star}$ \cite{bistritzer2011moire, lopes2007graphene}. 

Several conclusions can be directly inferred from the renormalized energies $E_{\ell}$. Let us first consider the simple case of only a mass perturbation, which opens a gap $\Delta_{\ell}^{\star}=2m_{\ell}^{\star}$ at both Dirac points. For arbitrary $\alpha$ and $m_{t}, m_{b}$, the gaps $\Delta_{t}^{\star}$ and $\Delta_{b}^{\star}$ are, in general, different. The gaps become equal when
\begin{equation}
\left(m_{t}-m_{b}\right)\left[1-3\alpha^{2}\left(1-\lambda^{2}\right)\right]=0.\label{eq:equi_mass}
\end{equation}
This is always satisfied if $m_{t}=m_{b}$, but also at the particular point where $3\alpha^{2}\left(1-\lambda^{2}\right)=1$, independently of the values of $m_{t}$ and $m_{b}$. The latter condition, met around the first magic angle where $3\alpha^{2}\sim 1$ and $v^{\star}\rightarrow 0$, provides a first manifestation of what we term \emph{moiré-driven equilibrium}, where distinct layer perturbations naturally flow toward a common value as a consequence of the moiré coupling. The equilibrium gap reads
\begin{equation}
\Delta_{\mathrm{eq}}^{\star}=\left(1-\lambda^{2}\right)\frac{\Delta_{t}+\Delta_{b}}{2},\label{eq:bandgap_renorm}
\end{equation}
where $\Delta_{\ell}=2m_{\ell}$ are the initial decoupled gaps. In the chiral model~\cite{Tarnopolsky2019Origin,Naumis2021Reduction} ($\lambda=0$), one has  $\Delta_{\mathrm{eq}}^{\star}=\left(\Delta_{t}+\Delta_{b}\right)/2$, reflecting a perfect equilibrium tendency. As $\lambda$ increases, the equilibrium gap decreases from the initial average $\left(\Delta_{t}+\Delta_{b}\right)/2$. In the rigid case $\lambda=1$, Eq.~\eqref{eq:bandgap_renorm} indicates that $\Delta_{\mathrm{eq}}^{\star}\rightarrow0$, but this reflects the breakdown of the first-order approximation \cite{SM}.

Similar conclusions can be drawn for the scalar perturbation that shifts the Dirac points in energy. The equilibrium condition is then given by
\begin{equation}
\left(V_{t}-V_{b}\right)\left[1-3\alpha^{2}\left(1+\lambda^{2}\right)\right]=0,\label{eq:equi_scalar}
\end{equation}
which is slightly different from the mass equilibrium condition due to the factor $\sim\left(1+\lambda^{2}\right)$. For nonzero $\lambda$, the shift equilibrium thus occurs at larger twist angles than the mass equilibrium. The equilibrium shift is given by
\begin{equation}
V_{\mathrm{eq}}^{\star}=\frac{V_{t}+V_{b}}{2},\label{eq:shift_renorm}
\end{equation}
and is independent of the ratio $\lambda$. The energy shift equilibrium reflects a second key manifestation of a moiré-driven equilibrium.

\begin{figure*}[t]
    \includegraphics[width=1\linewidth]{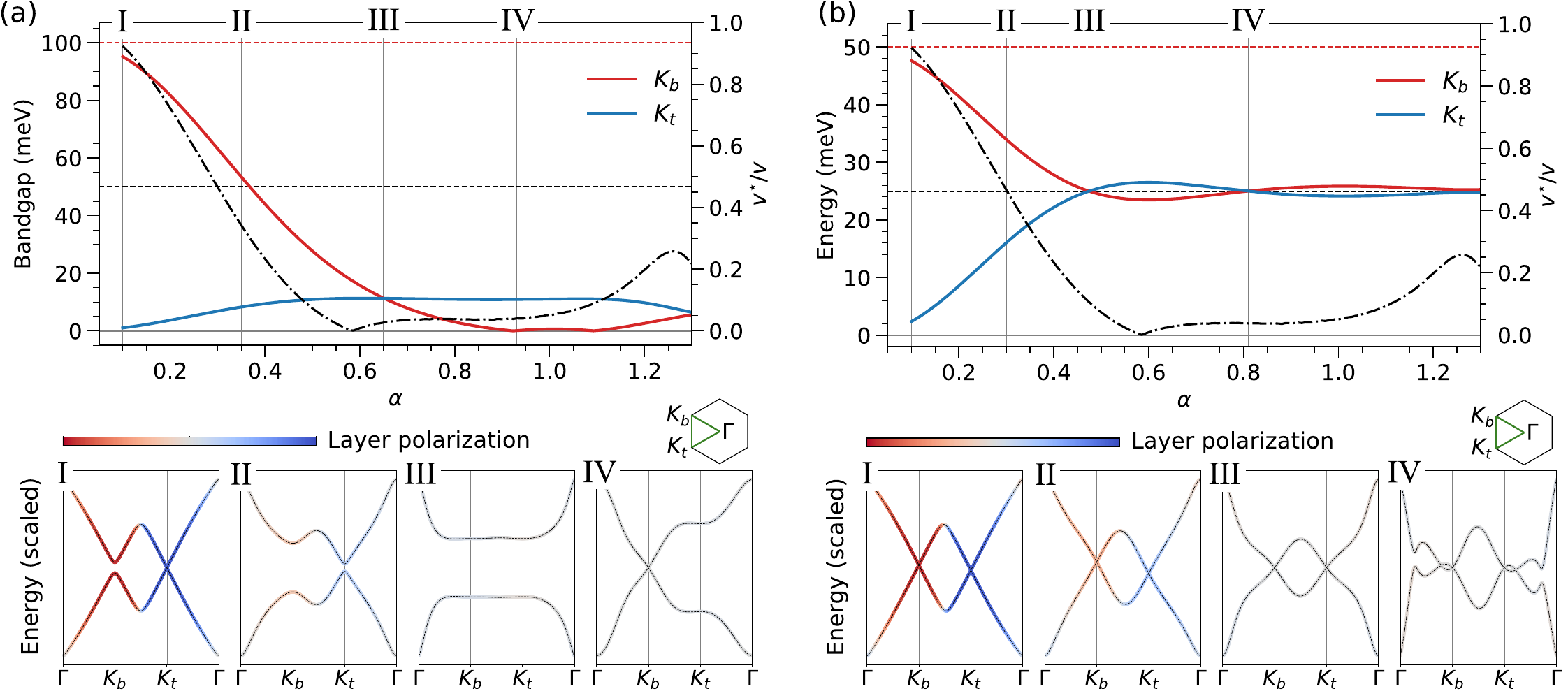}
    \caption{Moiré-driven equilibrium for (a) mass perturbation $P_{b}=\sigma_{z}m,P_{t}=0$ with $m=50\,\mathrm{meV}$ and (b) scalar perturbation $P_{b}=\sigma_{0}V,P_{t}=0$ with $V=50\,\mathrm{meV}$. In the decoupled system ($\alpha\rightarrow0$), the perturbation (a) opens a bandgap $\Delta_{b}=2m$ or (b) shift the Dirac point $E_b\rightarrow V$ at the perturbed $K_b$ point associated with the bottom layer. Top panels show the evolution of the bandgap and energy at the $K_b$ and $K_t$ points, as a function of the moiré coupling $\alpha$. The moiré-driven equilibrium takes place at (a) $\alpha=0.65$ and (b) $\alpha=0.475$. In both perturbations, the black dot-dashed curve shows the renormalized Fermi velocity ratio $v^{\star}/v$ of the unperturbed system (right axis), which vanishes at the first magic angle $\alpha\sim0.58$. Bottom panels show the moiré narrow bands at the points I, II, III and IV indicated in the top panels, corresponding to (a) $\alpha=0.1,0.35,0.65,0.93$ and (b) $\alpha=0.1,0.3,0.475,0.81$, respectively. The energy is scaled for better visualization. The colormaps indicate the layer polarization at each $k$ point, from bottom (red) to top (blue) polarization.} \label{fig:mass_scalar_perturbation}
\end{figure*}

For a gauge perturbation, the momentum shift $\delta\mathbf{k}_{\ell}=\mathbf{A}_{\ell}^{\star}/\hbar v^{\star}$ of the Dirac points can reach an equilibrium if $\delta\mathbf{k}_{t}=\pm\delta\mathbf{k}_{b}$, which occurs when
\begin{equation}
\left(\mathbf{A}_{t}-\mathbf{A}_{b}\right)\left(1\pm3\alpha^{2}\right)=0.
\end{equation}
This condition can only happen in the trivial case $\mathbf{A}_t=\mathbf{A}_b$ (with $\delta\mathbf{k}_{t}=\delta\mathbf{k}_{b}$ for all $\alpha$), or when $\delta\mathbf{k}_{t}=-\delta\mathbf{k}_{b}$ at the first magic angle where $3\alpha^{2}=1$. Note, however, that in the latter case the perturbation approximation no longer holds because both shifts $\delta\mathbf{k}_{\ell}$ diverge. 

The momentum-shift of the Dirac points implies the possibility of their collapse within the mBZ. The renormalized position of the Dirac points is given by $\mathbf{K}_{\ell}^{\star}=\mathbf{K}_{\ell}+\mathbf{A}_{\ell}^{\star}/\hbar v^{\star}$. The equilibrium condition $\mathbf{K}_{t}^{\star}=\mathbf{K}_{b}^{\star}$ takes place when
\begin{equation}
\mathbf{K}_{t}-\mathbf{K}_{b}=\hbar v\frac{1+3\alpha^{2}}{1-3\alpha^{2}}\left(\mathbf{A}_{b}-\mathbf{A}_{t}\right).\label{eq:DP_collapse}
\end{equation}
This condition depends on the twist angle both through the moiré coupling $\alpha$ and the difference $\mathbf{K}_{t}-\mathbf{K}_{b}$ which determines the length of the mBZ. Within this first-order result, the Dirac points collapse \textit{only} if the initial perturbation difference $\left(\mathbf{A}_{b}-\mathbf{A}_{t}\right)$ is along the same direction as $\left(\mathbf{K}_{t}-\mathbf{K}_{b}\right)$; if that is not the case, the Dirac points may become close but not actually collapse in momentum space. Remarkably, the summation of the renormalized Dirac points
\begin{equation}
\mathbf{K}_{t}^{\star}+\mathbf{K}_{b}^{\star}=\mathbf{K}_{t}+\mathbf{K}_{b}+\frac{\mathbf{A}_{t}+\mathbf{A}_{b}}{\hbar v},
\end{equation}
depends only on the initial perturbation condition and $\mathbf{K}_{t}+\mathbf{K}_{b}=2\cos\left(\theta/2\right)\mathbf{K}$. Since at low twist angles $\mathbf{K}_{t}+\mathbf{K}_{b}\approx2\mathbf{K}$, the collapse position $\mathbf{K}_{c}^{\star}=\mathbf{K}_{t}^{\star}=\mathbf{K}_{b}^{\star}$ is \emph{solely} determined by the initial perturbation:
\begin{equation}
\mathbf{K}_{c}^{\star}\approx\mathbf{K}+\frac{\mathbf{A}_{t}+\mathbf{A}_{b}}{2\hbar v}.\label{eq:Kc}
\end{equation}
Thus, the initial perturbation fully {\it determines} when and where the Dirac points collapse once the layers are coupled by the moiré potential. The momentum shift equilibrium, or the collapse of Dirac points, represents a third notable manifestation of a moiré-driven equilibrium.

\begin{figure*}[t]
    \includegraphics[width=1\linewidth]{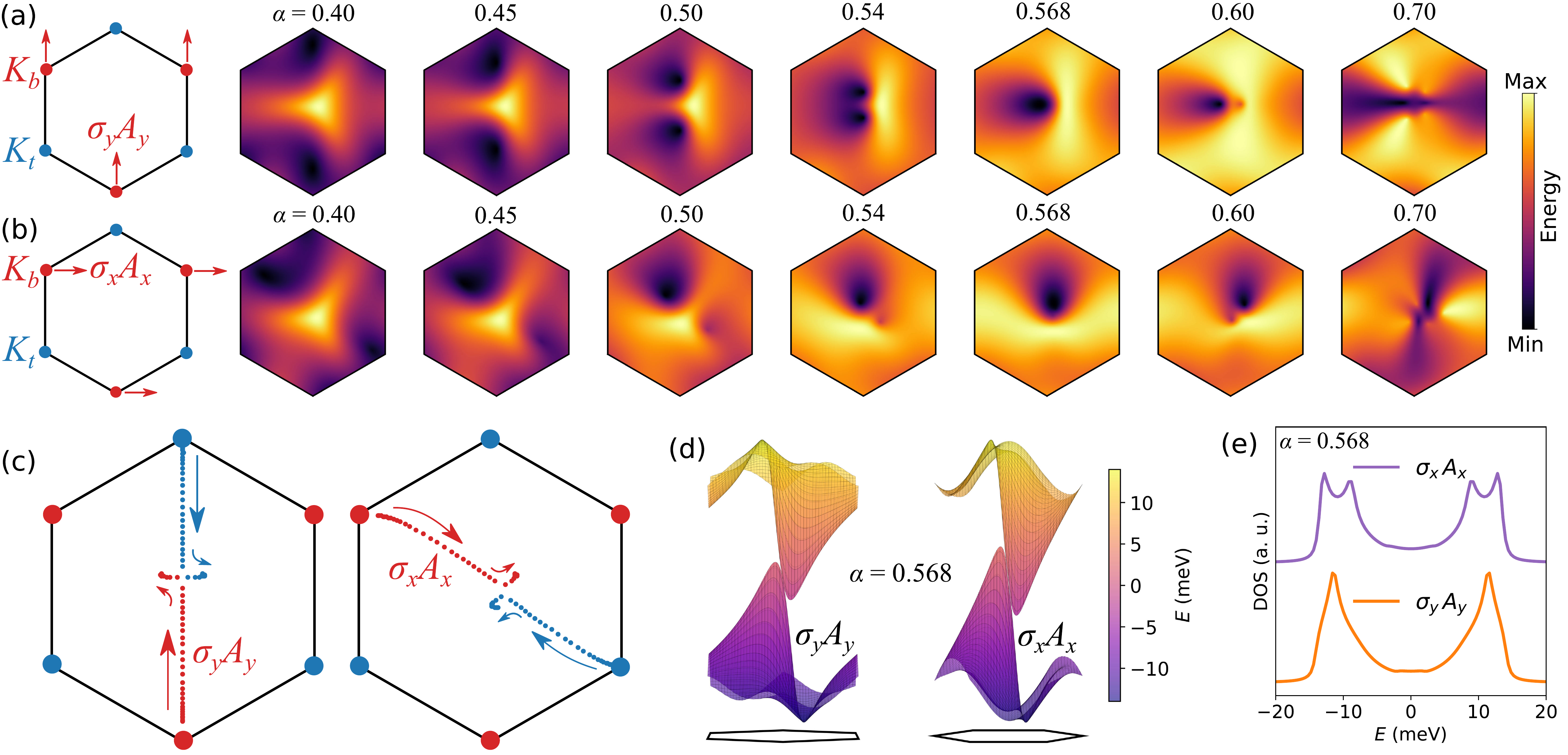}
    \caption{Moiré-driven equilibrium for a gauge perturbation (a) $P_{b}=\sigma_{y}A_{y}$ and (b) $P_{b}=\sigma_{x}A_{x}$, with $A_{x},A_{y}=0.03\,\mathrm{nm^{-1}}$. In both cases $P_{t}=0$, so in the decoupled system ($\alpha\rightarrow0$) the perturbation only shifts the Dirac points $K_b$ coming from the bottom layer. The top panels show the evolution of the top central moiré band at different moiré couplings $\alpha$, highlighting the collapse of the Dirac points (dark spots) as $\alpha$ increases. The collapse takes place around $\alpha\sim0.568$. Panel (c) shows the path of the $K_b$ and $K_t$ point as the moiré coupling increases, from $\alpha=0.3$ to $\alpha=0.7$ in steps of $0.01$; arrows indicate the direction of the path. (d) 3D plot of the two central moiré bands at the collapse point $\alpha=0.568$, with the moiré BZ shown in black below (the plots are rotated for better visualization). (e) Density of states at the collapse point, within the energy window of the narrow bands.}\label{fig:gauge}
\end{figure*}

\textit{Numerical results}— The previous first-order analysis points out that any perturbation will invariably tend to an equilibrium around the magic angle. We will now show that this behavior holds beyond first-order perturbation, as supported by numerical results of the full continuum model. Unless otherwise stated, we use $\hbar v/a=2.1354\,\mathrm{eV}$, $u'=0.0975\,\mathrm{eV}$ and $u/u'=0.8$ \cite{koshino2018maximally}.  

We first consider the simplest cases of a mass or a scalar perturbation on the bottom layer, so that $P_{b}=\sigma_{z}m$ or $P_{b}=\sigma_{0}V$, with $P_{t}=0$ in both cases. These perturbations initially (in the decoupled system) open a gap or shift the Dirac cone solely at the $K_b$ point associated with the bottom layer. Figure~\ref{fig:mass_scalar_perturbation} shows how the perturbations then evolve as the moiré coupling increases. In both cases, we observe a clear equilibrium tendency as one approaches the first (unperturbed) magic angle ($\alpha\sim0.58$). In line with Eqs.~\eqref{eq:equi_mass} and~\eqref{eq:equi_scalar}, the equilibrium point depends on the type of perturbation: for the mass perturbation, it occurs at a slightly lower twist angle (higher $\alpha$) than the first magic angle, whereas for the scalar perturbation it occurs at a slightly higher twist angle (lower $\alpha$). The overall equilibrium tendency, including the vanishing of the Fermi velocity, thus takes place around a restricted coupling regime $\alpha\sim0.55\pm0.1$.

Figure~\ref{fig:mass_scalar_perturbation} reveals other noteworthy features.  First, the equilibrium bandgap is lower than the initial average gap, i.e., $\Delta_{\mathrm{eq}}^{\star}<m$, in line with Eq.~\eqref{eq:bandgap_renorm} for $\lambda\neq0$. The equilibrium bandgap becomes equal to the initial gap ($\Delta_{\mathrm{eq}}^{\star}=m$) only in the chiral limit $\lambda=0$~\cite{SM}. By contrast, the equilibrium shift $V_{\mathrm{eq}}^{\star}=V/2$ is always half of the initial shift, regardless of $\lambda$, in line with Eq.~\eqref{eq:shift_renorm}. 
Second, beyond the first magic angle the effect of both perturbations tends to be reversed, in the sense that they become stronger in the unperturbed Dirac point ($K_t$). Hence, as $\alpha$ increases the bandgap and energy shift at $K_b$ can become lower than at $K_t$. For the scalar perturbation, this reverse effect follows an oscillatory behavior around the equilibrium value. For the mass perturbation, we remarkably find that the bandgap at the perturbed Dirac point can even close (e.g. at $\alpha\sim0.93$). As $\alpha$ continues to increase, one actually observes a highly nontrivial closing and reopening of the bandgap at both Dirac points~\cite{SM}, reflecting clear differences between the first magic angle and higher magic angles~\cite{crepel2025topologically,navarro2022first}.

The moiré-driven equilibrium under a mass perturbation is accompanied by a similar redistribution of the Berry curvature \cite{xiao2010berry}. Specifically, while at weak moiré couplings the Berry curvature is highly concentrated around the less gaped (unperturbed) Dirac point, at the equilibrium regime the Berry curvature becomes uniform at both Dirac points \cite{SM}. The closing and reopening of the bandgaps at stronger couplings further implies nontrivial changes in the topology of the central moiré bands.

Last panels in Fig.~\ref{fig:mass_scalar_perturbation} showcase the moiré narrow bands at selected values of the coupling $\alpha$. The layer polarization colormap at each $k$ point~\cite{SM} highlights that the equilibrium tendency is driven by the hybridization of the two layers induced by the moiré coupling. The equilibrium state occurs when the layers hybridize so that the initial perturbation is distributed equally between both Dirac points. This behavior implies that layer-polarized states \textit{cannot} occur around the magic angle, regardless of whether the system is perturbed or not \cite{cea2020band}. Note, however, that the bands can still be polarized in other flavors. For example, at the mass equilibrium point the top and bottom narrow bands are polarized in the A and B sublattices \cite{SM}.

Next, we consider the case of a gauge perturbation. Figure~\ref{fig:gauge} shows the results for two cases: $P_{b}=\sigma_{y}A_{y}$ and $P_{b}=\sigma_{x}A_{x}$, with again $P_{t}=0$. The density plots in panels (a) and (b) clearly show that the Dirac points (dark spots) tend to collapse at $\alpha\sim0.568$. The collapse path followed by each Dirac point depends on the initial perturbation. In the case $P_{b}=\sigma_{y}A_{y}$, the perturbation $\mathbf{A}_{b}\propto\mathbf{e}_{y}$ aligns with the difference between the twisted Dirac points, $\left(\mathbf{K}_{t}-\mathbf{K}_{b}\right)\propto k_{\theta}\mathbf{e}_{y}$, and the collapse path occurs along the $y$ axis, in agreement with the first-order prediction of Eq.~\eqref{eq:DP_collapse}. In the other case, $P_b=\sigma_{x}A_{x}$, the perturbation $\mathbf{A}_{b}\propto\mathbf{e}_{x}$ is not aligned with the difference $\mathbf{K}_{t}-\mathbf{K}_{b}$, and the collapse path is not along the perturbation direction. This latter behavior is markedly different from the first-order perturbation result, which predicts that the shift of the Dirac points takes place along the direction of the perturbation.

In general, the collision path is along the perturbation direction if it aligns with the transfer vectors connecting opposite Dirac points~\cite{SM}. This can be understood by noting that the collision path should minimize the energy associated with shifting the Dirac points, which translates to minimum deviations from the perturbation direction. Thus, the Dirac points collapse along perturbation directions that connect high-symmetry points, whereas for other directions the collapse path necessarily bents with respect to the perturbation.

The moiré coupling can also shift the Dirac points in energy, even in the absence of a perpendicular electric field. This effect is not captured by the first-order energy in Eq.~\eqref{eq:El_pert} because the perturbation is performed around the zero-energy position of the unperturbed Dirac points. As seen in Figs.~\ref{fig:gauge}(a) and (b), the energetic shift is found to be pronounced when the perturbation $\mathbf{A}_{b}$ is not aligned with the difference $\left(\mathbf{K}_{t}-\mathbf{K}_{b}\right)$. The energetic shift of the Dirac points is further reflected in Fig.~\ref{fig:gauge}(d), which shows a three-dimensional plot of the narrow bands and the density of states at the collapse point $\alpha=0.568$. In the case $P_{b}=\sigma_{y}A_{y}$, the two Dirac points remain almost at the same energy as the collapse is approached, whereas in the case $P_{b}=\sigma_{x}A_{x}$ the momentum-space collapse occurs at different energies. After the collapse, however, the Dirac points shift in energy in both cases. As $\alpha$ increases further, both the momentum and energy shift of the Dirac points exhibit a nontrivial oscillatory behavior, similar to that observed in Fig.~\ref{fig:mass_scalar_perturbation}, reflecting once again that the first magic angle is special~\cite{crepel2025topologically, navarro2022first}.

\textit{Connection to experiments}— In most realistic scenarios, perturbations that modify intralayer properties also affect the interlayer coupling, raising the question of whether the moiré-driven equilibrium remains robust when the moiré potential itself is perturbed. A representative example is twisted bilayer graphene on hexagonal boron nitride (hBN)~\cite{zhang2019twisted, cea2020band, lin2020symmetry, shi2021moire}. In addition to breaking inversion symmetry and inducing a mass term, hBN generates effective periodic potentials~\cite{mucha2013heterostructures, Kindermann2012Zero, Wallbank2013Generic, SanJose2014Electronic, SanJose2014Spontaneous}. Since hBN shares the same hexagonal symmetry as graphene, these additional potentials preserve the $C_{3}$ symmetry and act in close analogy to the unperturbed moiré potential. One therefore expects the \emph{moiré-driven equilibrium} to persist even in the presence of such periodic fields. This expectation is supported by atomistic and continuum calculations by Long \emph{et al.}~\cite{long2022atomistic, long2023electronic}, who showed that an unequal mass perturbation flows toward equilibrium in both the gap and the charge distribution near the magic angle. 

Gauge perturbations arise most naturally in strained samples~\cite{huder2018electronic, kazmierczak2021strain, mesple2021heterostrain}. In general, strain introduces both gauge and scalar potentials and simultaneously modifies the moiré potential through changes in the moiré vectors~\cite{escudero2024designing, huder2018electronic, bi2019designing}. In Ref.~\cite{SM}, we examined several strain configurations within a strain-extended continuum model~\cite{Pantaleon2022Interaction, escudero2024designing} and found that the resulting behavior still reflects a clear moiré-driven equilibrium, with the Dirac points shifting toward a collapse as the moiré coupling increases.

These results provide a natural framework to interpret recent experiments on strained TBG. In particular, Refs.~\cite{yu2024twist, carrasco2025twistraintronics} investigated the effects of strain in TBG and found that the observed electronic features near the magic angle can be reproduced assuming strain applied to a single layer. Notably, in Ref.~\cite{yu2024twist}, a combined experimental and theoretical analysis of samples with spatially varying twist and strain demonstrated that near the magic angle the electronic response is, essentially, identical whether strain is present in one layer or in both layers. This apparent insensitivity follows naturally from the \emph{moiré-driven equilibrium} identified here, which effectively masks the layer-resolved origin of the perturbation under strong moiré coupling.

\textit{Conclusions}— We have shown that the magic-angle regime in twisted bilayer graphene is part of a broader phenomenon that we term \emph{moiré-driven equilibrium}, whereby perturbations initially acting on individual layers become redistributed once the layers are strongly coupled by the moiré potential. Using first-order perturbation theory supported by extensive numerical calculations, we demonstrated that this equilibrium mechanism governs mass, scalar, and gauge perturbations. In particular, it leads to equal bandgaps for mass terms, equal energy shifts for scalar potentials, and equal momentum-shifts for gauge perturbations that can culminate in a collapse of Dirac points within the moiré Brillouin zone. Beyond the first magic angle, the redistribution of perturbations persists and follows a nontrivial oscillatory behavior, highlighting qualitative differences between the first and higher magic angles. By connecting our results to recent experiments, we showed that the moiré-driven equilibrium remains robust even when the moiré potential itself is perturbed, providing a natural explanation for the observed masking of layer-resolved properties near the magic angle. Since this equilibrium mechanism arises from strong interlayer hybridization induced by the moiré coupling, we expect a tendency toward moiré-driven equilibrium to be a generic feature of moiré materials.

\textit{Acknowledgments}— We thank Saul Herrera, Gerardo G. Naumis and Tommaso Cea for discussions. We acknowledged support from the “Severo Ochoa” Programme for Centres of Excellence in R\&D (CEX2020-001039-S/AEI/10.13039/501100011033) financed by MICIU/AEI/10.13039/501100011033 and from NOVMOMAT, Grant PID2022-142162NB-I00 funded by MCIN/AEI/ 10.13039/501100011033 and, by "ERDF A way of making Europe". F.E. acknowledges support funding from the European Union's Horizon 2020 research and innovation programme under the Marie Skłodowska-Curie grant agreement No 101210351. Z.Z. acknowledges support funding from the European Union's Horizon 2020 research and innovation programme under the Marie Skłodowska-Curie grant agreement No 101034431.  P.A.P acknowledges funding by Grant No.\ JSF-24-05-0002 of the Julian Schwinger Foundation for Physics Research. F.G. acknowledges the support from the  Department of Education of the Basque Government through the project No. \verb|PIBA\2023\1\0007(STRAINER)|.

\let\oldaddcontentsline\addcontentsline
\renewcommand{\addcontentsline}[3]{}

\bibliography{References.bib}


\let\addcontentsline\oldaddcontentsline

\clearpage
\onecolumngrid

\setcounter{equation}{0}
\setcounter{figure}{0}
\setcounter{table}{0}
\setcounter{page}{1}
\renewcommand{\theequation}{S\arabic{equation}}
\renewcommand{\thefigure}{S\arabic{figure}}
\setcounter{secnumdepth}{3}

\begin{center}
{\Large\emph{Supplemental Materials for}:\\
Moiré-driven equilibrium of perturbations in moiré systems}{\Large\par}
\par\end{center}

\begin{center}
Federico Escudero, Zhen Zhan, Pierre A. Pantale\'on, and Francisco Guinea
\par\end{center}

\tableofcontents
\let\oldaddcontentsline\addcontentsline

\section{First-order perturbation theory}

Truncating the continuum model Hamiltonian of TBG up to the first shell implies taking into account only the three leading-order couplings between electrons in different layers \cite{bistritzer2011moire}. For the moiré potential given by Eq. (3) in the main text \cite{koshino2015interlayer, koshino2018maximally}, the three couplings correspond to momenta exchange \cite{bistritzer2011moire}
\begin{align}
\mathbf{q}_{1} & =\mathbf{K}_{b}-\mathbf{K}_{t}=-\frac{2\mathbf{G}_{1}+\mathbf{G}_{2}}{3},\\
\mathbf{q}_{2} & =\mathbf{q}_{1}+\mathbf{G}_{1},\\
\mathbf{q}_{3} & =\mathbf{q}_{1}+\mathbf{G}_{1}+\mathbf{G}_{2},\label{eq:q_transer}
\end{align}
where $\mathbf{G}_{1}$ and $\mathbf{G}_{2}$ are the moiré vectors. For lattice vectors $\mathbf{a}_{1}=a\left(1,0\right)$, $\mathbf{a}_{2}=a\left(1/2,\sqrt{3}/2\right)$ (with $a\approx2.46\,\textrm{\AA}$ in graphene), and a symmetric twist configuration $\pm\theta/2$, one has $\mathbf{q}_{1}=k_{\theta}\left(0,1\right)$, $\mathbf{q}_{2}=R\left(2\pi/3\right)\mathbf{q}_{1}$ and $\mathbf{q}_{3}=R\left(4\pi/3\right)\mathbf{q}_{1}$, where $k_{\theta}=8\pi\sin\left(\theta/2\right)/3a$ is the length of the moiré Brillouin zone.

To study the perturbed energies around both Dirac points, we follow Ref. \cite{bistritzer2011moire} and truncate the TBG Hamiltonian around momenta $\mathbf{k}=\mathbf{K}_{t}$ and $\mathbf{k}=\mathbf{K}_{b}$. Taking into account the generic perturbations $P_{\ell}$ in each layer (see main text), the two first-shell truncated Hamiltonian read
\begin{equation}
h_{t}=\left(\begin{array}{cccc}
-\hbar v\boldsymbol{\sigma}\cdot\mathbf{k}_{t}+P_{t} & U_{1} & U_{2} & U_{3}\\
U_{1} & -\hbar v\boldsymbol{\sigma}\cdot\left(\mathbf{k}_{t}-\mathbf{q}_{1}\right)+P_{b} & 0 & 0\\
U_{2} & 0 & -\hbar v\boldsymbol{\sigma}\cdot\left(\mathbf{k}_{t}-\mathbf{q}_{2}\right)+P_{b} & 0\\
U_{3} & 0 & 0 & -\hbar v\boldsymbol{\sigma}\cdot\left(\mathbf{k}_{t}-\mathbf{q}_{3}\right)+P_{b}
\end{array}\right),
\end{equation}
\begin{equation}
h_{b}=\left(\begin{array}{cccc}
-\hbar v\boldsymbol{\sigma}\cdot\mathbf{k}_{b}+P_{b} & U_{1} & U_{2} & U_{3}\\
U_{1} & -\hbar v\boldsymbol{\sigma}\cdot\left(\mathbf{k}_{b}+\mathbf{q}_{1}\right)+P_{t} & 0 & 0\\
U_{2} & 0 & -\hbar v\boldsymbol{\sigma}\cdot\left(\mathbf{k}_{b}+\mathbf{q}_{2}\right)+P_{t} & 0\\
U_{3} & 0 & 0 & -\hbar v\boldsymbol{\sigma}\cdot\left(\mathbf{k}_{b}+\mathbf{q}_{3}\right)+P_{t}
\end{array}\right),
\end{equation}
where $\mathbf{k}_{\ell}=\mathbf{k}-\mathbf{K}_{\ell}$ is the momentum measured with respect to the Dirac point $\mathbf{K}_{\ell}$. Note that we have used that $U_{j}^{\dagger}=U_{j}$. The truncated $8\times8$ Hamiltonian $h_{\ell}$ gives 8 bands that capture the low-energy spectra of the full continuum model around the Dirac points and up to the first-magic angle. Importantly, the Hamiltonian $h_{\ell}$ can capture the emergence of flat bands and the main effects of the perturbations $P_{\ell}$.

To obtain the first order correction to the energies around the Dirac points, we separate
\begin{equation}
h_{\ell}=h_{\ell,0}+h'_{\ell},
\end{equation}
where $h_{\ell,0}$ and $h'_{\ell}$ are the unperturbed and perturbed Hamiltonian
\begin{align}
h_{\ell,0} & =\left(\begin{array}{cccc}
0 & U_{1} & U_{2} & U_{3}\\
U_{1} & \pm\hbar v\boldsymbol{\sigma}\cdot\mathbf{q}_{1} & 0 & 0\\
U_{2} & 0 & \pm\hbar v\boldsymbol{\sigma}\cdot\mathbf{q}_{2} & 0\\
U_{3} & 0 & 0 & \pm\hbar v\boldsymbol{\sigma}\cdot\mathbf{q}_{3}
\end{array}\right),\\
h'_{\ell} & =\left(\begin{array}{cccc}
-\hbar v\boldsymbol{\sigma}\cdot\mathbf{k}_{\ell}+P_{\ell} & 0 & 0 & 0\\
0 & -\hbar v\boldsymbol{\sigma}\cdot\mathbf{k}_{\ell}+P_{\ell'} & 0 & 0\\
0 & 0 & -\hbar v\boldsymbol{\sigma}\cdot\mathbf{k}_{\ell}+P_{\ell'} & 0\\
0 & 0 & 0 & -\hbar v\boldsymbol{\sigma}\cdot\mathbf{k}_{\ell}+P_{\ell'}
\end{array}\right).
\end{align}
In $h_{\ell,0}$, $+$ ($-$) is for the top (bottom) layer and $\ell'\neq\ell$ refers to the other layer. The first order correction to the energy reads
\begin{align}
E_{\ell} & =\frac{\left\langle \Psi_{0}\left|h'_{\ell}\right|\Psi_{0}\right\rangle }{\left\langle \Psi_{0}\mid\Psi_{0}\right\rangle },
\end{align}
where $\left|\Psi_{0}\right\rangle $ are the eigenstates of $h_{\ell,0}$, with solution $E_{\ell,0}=0$ and \cite{bistritzer2011moire}
\begin{equation}
\Psi_{0}=\left(\begin{array}{c}
\psi_{0}\\
\mp h_{1}^{-1}U_{1}\psi_{0}\\
\mp h_{2}^{-1}U_{2}\psi_{0}\\
\mp h_{3}^{-1}U_{3}\psi_{0}
\end{array}\right),
\end{equation}
where $h_{j}=\hbar v\boldsymbol{\sigma}\cdot\mathbf{q}_{j}$ ($j=1,2,3$) and $\psi_{0}$ is a $2\times1$ spinor. Using the property $\sum_{j=1}^{3}U_{j}h_{j}^{-1}U_{j}=0$ and taking $\left|\psi_{0}\right|^{2}=1$ yields \cite{bistritzer2011moire}
\begin{equation}
\left\langle \Psi_{0}\mid\Psi_{0}\right\rangle =1+3\alpha^{2}\left(1+\lambda^{2}\right),
\end{equation}
where, as in the main text, 
\begin{equation}
\alpha=\frac{u'}{\hbar vk_{\theta}},\quad\lambda=\frac{u}{u'}.
\end{equation}

Next, we have 
\begin{align}
\left\langle \Psi_{0}\left|h'_{\ell}\right|\Psi_{0}\right\rangle  & =-\psi_{0}^{\dagger}\left(\hbar v\boldsymbol{\sigma}\cdot\mathbf{k}_{\ell}\right)\psi_{0}-\psi_{0}^{\dagger}\sum_{j=1}^{3}U_{j}h_{j}^{-1}\left(\hbar v\boldsymbol{\sigma}\cdot\mathbf{k}_{\ell}\right)h_{j}^{-1}U_{j}\psi_{0}\nonumber\\
 & +\psi_{0}^{\dagger}P_{\ell}\psi_{0}+\psi_{0}^{\dagger}\sum_{j=1}^{3}U_{j}h_{j}^{-1}P_{\ell'}h_{j}^{-1}U_{j}\psi_{0},\label{eq:pert_1}
\end{align}
where, again, $\ell'\neq\ell$ refers to the other layer and we have used that $\left(h_{j}^{-1}\right)^{\dagger}=h_{j}^{-1}$. The second term on the r.h.s of Eq. \eqref{eq:pert_1} is the same as without perturbation \cite{bistritzer2011moire}
\begin{equation}
\sum_{j=1}^{3}U_{j}h_{j}^{-1}\left(\hbar v\boldsymbol{\sigma}\cdot\mathbf{k}_{\ell}\right)h_{j}^{-1}U_{j}=-3\alpha^{2}\hbar v\boldsymbol{\sigma}\cdot\mathbf{k}_{\ell}.
\end{equation}
The first-order corrected energies are thus given by $E_{\ell}=\psi_{0}^{\dagger}H_{\ell}^{\star}\psi$, where $H_{\ell}^{\star}$ is the $2\times2$ perturbed Hamiltonian given by Eq. (5) in the main text:
\begin{align}
H_{\ell}^{\star} & =\frac{-\hbar v\boldsymbol{\sigma}\cdot\mathbf{k}_{\ell}\left(1-3\alpha^{2}\right)+P_{\ell}+\sum_{j=1}^{3}U_{j}h_{j}^{-1}P_{\ell'}h_{j}^{-1}U_{j}}{1+3\alpha^{2}\left(1+\lambda^{2}\right)}\nonumber \\
 & =-\hbar v^{\star}\boldsymbol{\sigma}\cdot\left(\mathbf{k}-\mathbf{K}_{\ell}\right)+P_{\ell}^{\star},\label{eq:Hl_SM}
\end{align}
where
\begin{align}
v^{\star} & =v\frac{1-3\alpha^{2}}{1+3\alpha^{2}\left(1+\lambda^{2}\right)},\\
P_{\ell}^{\star} & =\frac{P_{\ell}+\sum_{j=1}^{3}U_{j}h_{j}^{-1}P_{\ell'}h_{j}^{-1}U_{j}}{1+3\alpha^{2}\left(1+\lambda^{2}\right)},
\end{align}
are the renormalized Fermi velocity and perturbation. Eq. \eqref{eq:Hl_SM} is valid for any perturbation $P_{\ell}$. For the scalar, mass and gauge perturbation
\begin{equation}
P_{\ell}=\sigma_{0}V_{\ell}+\boldsymbol{\sigma}\cdot\mathbf{A}_{\ell}+\sigma_{z}m_{\ell},
\end{equation}
we obtain
\begin{align}
\sum_{j=1}^{3}U_{j}h_{j}^{-1}\left(\sigma_{0}V_{\ell}\right)h_{j}^{-1}U_{j} & =3\alpha^{2}\sigma_{0}V_{\ell}\left(1+\lambda^{2}\right),\\
\sum_{j=1}^{3}U_{j}h_{j}^{-1}\left(\boldsymbol{\sigma}\cdot\mathbf{A}_{\ell}\right)h_{j}^{-1}U_{j} & =-3\alpha^{2}\boldsymbol{\sigma}\cdot\mathbf{A}_{\ell},\\
\sum_{j=1}^{3}U_{j}h_{j}^{-1}\left(\sigma_{z}m_{\ell}\right)h_{j}^{-1}U_{j} & =3\alpha^{2}\sigma_{z}m_{\ell}\left(1-\lambda^{2}\right).
\end{align}
The perturbed Hamiltonian then reads
\begin{align}
H_{\ell}^{\star} & =-\hbar v^{\star}\boldsymbol{\sigma}\cdot\left(\mathbf{k}-\mathbf{K}_{\ell}\right)+\sigma_0V_{\ell}^{\star}+\boldsymbol{\sigma}\cdot\mathbf{A}_{\ell}^{\star}+\sigma_{z}m_{\ell}^{\star},
\end{align}
where $V_{\ell}^{\star},\mathbf{A}_{\ell}^{\star}$ and $m_{\ell}^{\star}$
are the renormalized perturbations
\begin{align}
V_{\ell}^{\star} & =\frac{V_{\ell}+V_{\ell'}3\alpha^{2}\left(1+\lambda^{2}\right)}{1+3\alpha^{2}\left(1+\lambda^{2}\right)},\\
m_{\ell}^{\star} & =\frac{m_{\ell}+m_{\ell'}3\alpha^{2}\left(1-\lambda^{2}\right)}{1+3\alpha^{2}\left(1+\lambda^{2}\right)},\\
\mathbf{A}_{\ell}^{\star} & =\frac{\mathbf{A}_{\ell}-\mathbf{A}_{\ell'}3\alpha^{2}}{1+3\alpha^{2}\left(1+\lambda^{2}\right)}.\label{eq:A_shift}
\end{align}
The corresponding energies are given by Eq. (7) in the main text:
\begin{equation}
E_{\ell}=V_{\ell}^{\star}\pm\sqrt{m_{\ell}^{\star2}+\left|\hbar v^{\star}\left(\mathbf{k}-\mathbf{K}_{\ell}\right)-\mathbf{A}_{\ell}^{\star}\right|^{2}}.
\end{equation}
Although the perturbations effect remain the same (namely, shifting the Dirac cones in energy and momentum, and opening a gap), their magnitude is crucially renormalized due to the moiré coupling. This result is analogous to the pristine first-order result \cite{lopes2007graphene, bistritzer2011moire}, in the sense that the dispersion around the Dirac points remains linear but with a renormalized Fermi velocity.

\section{Polarization operators}

We define the layer polarization operator $\hat{\mathcal{P}_{L}}$ by the relations
\begin{align}
\hat{\mathcal{P}_{L}}\left|\psi_{t}\right\rangle  & =\left|\psi_{t}\right\rangle ,\\
\hat{\mathcal{P}_{L}}\left|\psi_{b}\right\rangle  & =-\left|\psi_{b}\right\rangle ,
\end{align}
where $\left|\psi_{t,b}\right\rangle $ are projected states in the top and bottom layers. In TBG, the Bloch states for an electron with momentum $\mathbf{k}$ in the moiré band $n$ (for a particular spin/valley flavor) read \cite{koshino2018maximally}
\begin{equation}
\left|\psi_{n,\mathbf{k},\ell,i}\right\rangle =\sum_{\mathbf{G}}u_{n,\mathbf{k},\ell,i}\left(\mathbf{G}\right)\left|\mathbf{k}+\mathbf{G}\right\rangle ,
\end{equation}
where $\mathbf{G}$ are moiré vectors, $\ell=t,b$ is the layer index, $i=A,B$ is the sublattice index, $\left|\mathbf{k}+\mathbf{G}\right\rangle $ are plane waves states, and $u_{n,\mathbf{k},\ell,i}\left(\mathbf{G}\right)$ are Fourier coefficients normalized according to 
\begin{equation}
\sum_{\mathbf{G},\ell,i}u_{n,\mathbf{k},\ell,i}^{*}\left(\mathbf{G}\right)u_{m,\mathbf{k},\ell,i}\left(\mathbf{G}\right)=\delta_{nm},\label{eq:norm_SM}
\end{equation}
which ensures that the total wave function is normalized within a moiré unit cell:
\begin{equation}
\sum_{\ell,i}\int_{\textrm{moiré unit cell}}d\mathbf{r}\left|\psi_{n,\mathbf{k},\ell,i}\left(\mathbf{r}\right)\right|^{2}=1.
\end{equation}
The polarization operator acting on $\left|\psi_{n,\mathbf{k},\ell,i}\right\rangle $ keeps or changes the sign depending on whether it corresponds to a top or bottom layer state. The layer polarization shown in Figure 2 of the main text corresponds to the expectation value 
\begin{align}
\mathcal{P}_{L} & =\sum_{\ell,i}\left\langle \psi_{n,\mathbf{k},\ell,i}\right|\hat{\mathcal{P}_{L}}\left|\psi_{n,\mathbf{k},\ell,i}\right\rangle \nonumber \\
 & =\sum_{\mathbf{G},i}\left(\left|u_{n,\mathbf{k},t,i}\left(\mathbf{G}\right)\right|^{2}-\left|u_{n,\mathbf{k},b,i}\left(\mathbf{G}\right)\right|^{2}\right).\label{eq:layer_pol}
\end{align}
Thus, if a state is totally polarized in the top layer, $u_{n,\mathbf{k},b,i}\left(\mathbf{G}\right)=0$ and, using Eq. \eqref{eq:norm_SM}, $\mathcal{P}_{L}=1$. Conversely, if a state is totally polarized in the bottom layer, then $u_{n,\mathbf{k},t,i}\left(\mathbf{G}\right)=0$ and $\mathcal{P}_{L}=-1$. Using the normalization condition of Eq. \eqref{eq:norm_SM} we can write
\begin{equation}
\mathcal{P}_{L}=2\sum_{\mathbf{G},i}\left|u_{n,\mathbf{k},t,i}\left(\mathbf{G}\right)\right|^{2}-1=1-2\sum_{\mathbf{G},i}\left|u_{n,\mathbf{k},b,i}\left(\mathbf{G}\right)\right|^{2}.
\end{equation}
In TBG, the $K_t$ and $K_b$ moiré Dirac points come from the top and bottom Dirac cones. Consequently, at low moiré couplings ($\alpha\ll1$) the states close to $K_t$ ($K_b$) have mostly a top (bottom) layer polarization; see Figure 2 in the main text. At strong moiré couplings, however, the layers are strongly hybridized and the polarization becomes uniform ($\mathcal{P}_{L} \sim0$) over the mBZ. 

We can similarly define a sublattice polarization operator $\hat{\mathcal{P}_{S}}$ by the relations
\begin{align}
\hat{\mathcal{P}_{S}}\left|\psi_{A}\right\rangle  & =\left|\psi_{A}\right\rangle ,\\
\hat{\mathcal{P}_{S}}\left|\psi_{B}\right\rangle  & =-\left|\psi_{B}\right\rangle ,
\end{align}
where now $\left|\psi_{A,B}\right\rangle $ are projected states in the $A$ and $B$ sublattices. The corresponding expectation value for the TBG Bloch states reads
\begin{align}
\mathcal{P}_{S} & =\sum_{\ell,i}\left\langle \psi_{n,\mathbf{k},\ell,i}\right|\hat{\mathcal{P}_{S}}\left|\psi_{n,\mathbf{k},\ell,i}\right\rangle \nonumber \\
 & =\sum_{\mathbf{G},\ell}\left(\left|u_{n,\mathbf{k},\ell,A}\left(\mathbf{G}\right)\right|^{2}-\left|u_{n,\mathbf{k},\ell,B}\left(\mathbf{G}\right)\right|^{2}\right),
\end{align}
which is analog to Eq. \eqref{eq:layer_pol} but with the layer and sublattice indices interchanged. 

\section{Strain-extended continuum model}\label{app:strain}

The continuum model of TBG can be extended to take into account the effect of strain \cite{bi2019designing}. The strain deforms the underlying honeycomb lattices and thus change the moiré pattern of the system \cite{escudero2024designing}. In general, the strain modifies the continuum model of TBG by: (i) changing the moiré vectors and therefore the moiré potential (see Eq. (2) in the main text); (ii) introducing scalar and gauge potentials coming from the change in the intralayer onsite and hopping energies \cite{Suzuura2002Phonons}. The strain-extended continuum model, for the $\xi=+$ valley, takes the form \cite{bi2019designing}
\begin{equation}
H=\left(\begin{array}{cc}
H_{t}+\mathcal{S}_{t} & U^{\dagger}\\
U & H_{b}+\mathcal{S}_{b}
\end{array}\right).
\end{equation}
The Dirac Hamiltonian are given by $\ensuremath{H_{\ell}=-\hbar v\boldsymbol{\sigma}\cdot\left(\mathbf{k}-\mathbf{K}_{\ell}\right)}$,
where now $\mathbf{K}_{\ell}$ are the twisted and strained Dirac points \cite{escudero2025geometrical}
\begin{equation}
\mathbf{K}_{\ell}=\left(\mathbb{I}-\mathcal{E}_{\ell}\right)R\left(\theta_{\ell}\right)\mathbf{K},
\end{equation}
where $\mathcal{E}_{\ell}$ is the strain tensor and $\mathbf{K}$ is the Dirac point of the honeycomb lattice. The strain-induced fields $\mathcal{S}_{\ell}$ read \cite{Suzuura2002Phonons, escudero2024designing}
\begin{equation}
\mathcal{S}_{\ell}=\sigma_0V_{\ell}-\hbar v\boldsymbol{\sigma}\cdot\mathbf{A}_{\ell},
\end{equation}
where $V_{\ell}$ and $\mathbf{A}_{\ell}$ are the scalar and gauge potentials
\begin{align}
V_{\ell} & =g\left(\epsilon_{xx}^{\ell}+\epsilon_{yy}^{\ell}\right),\\
\mathbf{A}_{\ell} & =\frac{\sqrt{3}}{2a}\beta\left(\epsilon_{xx}^{\ell}-\epsilon_{yy}^{\ell},-2\epsilon_{xy}^{\ell}\right),
\end{align}
where $\epsilon_{ij}^{\ell}$ ($i,j=x,y$) are the components of the strain tensor $\mathcal{E}_{\ell}$. For graphene we take $g=4$ eV and $\beta=3.14$. For simplicity, we continue to neglect the rotation and strain effect on the Pauli matrices in both $H_{\ell}$ and $S_{\ell}$ (they do not affect our conclusions). 

The moiré potential $U$ has the same leading order Fourier expansion as in TBG,
\begin{equation}
U\left(\mathbf{r}\right)=U_{1}+U_{2}e^{i\mathbf{G}_{1}\cdot\mathbf{r}}+U_{3}e^{i\left(\mathbf{G}_{1}+\mathbf{G}_{2}\right)\cdot\mathbf{r}},
\end{equation}
where $U_{j}$ are the matrices given by Eq. (3) in the main text
\begin{equation}
U_{j}=\left(\begin{array}{cc}
u & u'e^{-i\phi_{j}}\\
u'e^{i\phi_{j}} & u
\end{array}\right),
\end{equation}
with $\phi_{j}=\left(j-1\right)2\pi/3$. Although the matrices $U_{j}$ remain the same, the strain modifies the effect of the moiré potential due to the change in the moiré vectors $\mathbf{G}_{i}$ ($i=1,2$), which under twist and strain read \cite{escudero2024designing, escudero2025geometrical}
\begin{equation}
\mathbf{G}_{i}=\left[\left(\mathbb{I}-\mathcal{E}_{b}\right)R\left(\theta_{b}\right)-\left(\mathbb{I}-\mathcal{E}_{t}\right)R\left(\theta_{t}\right)\right]\mathbf{b}_{i},
\end{equation}
where $\mathbf{b}_{i}$ are the reciprocal vectors of the honeycomb lattice. The strain can also (slightly) modify the effective AA and AB/BA hopping parameters $u$ and $u'$ \cite{escudero2025}. However, to keep things simple, for the numerical calculations we continue to use the parameters $u'=0.0975\,\mathrm{eV}$, $u=\lambda u'$ and $\hbar v/a=2.1354\,\mathrm{eV}$ (the specific hopping energies do not change the moiré-driven equilibrium behavior).

\section{Extended numerical results}

Here we present extended numerical results for different perturbations, as a function of the coupling strength $\alpha$ and the ratio $\lambda$. Overall, these results reflect and support a global tendency towards a \textit{moiré-driven equilibrium}, regardless of the perturbation type and strength.

\subsection{Mass perturbation}

\begin{figure*}[t]
    \includegraphics[width=1\linewidth]{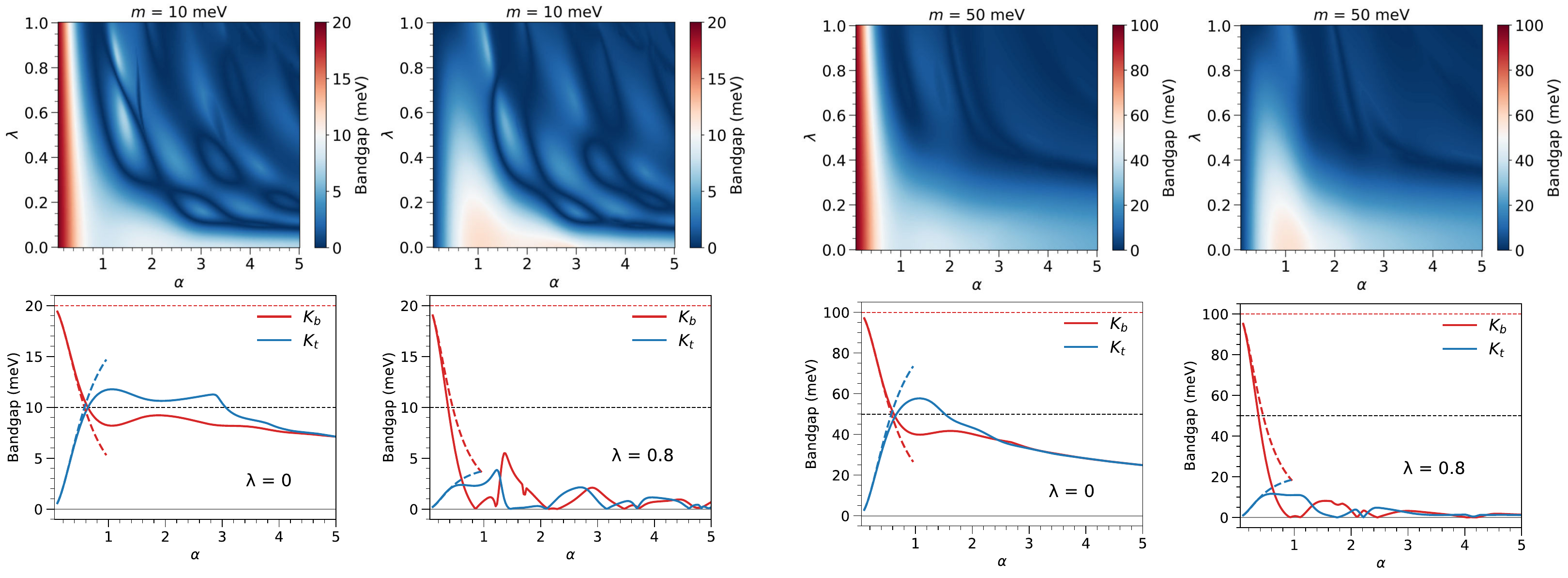}
    \caption{Bandgap evolution for a mass perturbation $P_{b}=\sigma_{z}m,P_{t}=0$, with different mass magnitudes. For each $m$, the top two panels show density plots of the bandgap at the perturbed bottom layer (left, $K_b$ point) and unperturbed top layer (right, $K_t$ point), as a function of the moiré coupling strength $\alpha$ and ratio $\lambda$. Bottom panels show, for each $m$, the bandgap at $K_b$ and $K_t$, as a function of $\alpha$ for $\lambda=0$ (left) and $\lambda=0.8$ (right). The horizontal red dashed-line indicates the decoupled bandgap $\Delta_b=2m$ at $K_b$, while the black dashed-line indicates the middle point $\Delta=m$. Dashed curves correspond to the first-order perturbation result given by Eq. (9) in the main text.
 }\label{fig:bandgap_SM}
\end{figure*}

Figure \ref{fig:bandgap_SM} shows numerical results for a mass perturbation $P_{b}=\sigma_{z}m$ in the bottom layer ($K_b$ point) with different values of $m$, and no perturbation in the top layer ($P_{t}=0$, $K_t$ point). For each mass, the top panels show a density map of the bandgap at the perturbed ($K_b$ point) and unperturbed ($K_t$ point) layers, while the bottom panel show the bandgap at $\lambda=0$ (chiral limit) and $\lambda=0.8$. 

Clearly, in all cases there is a mass equilibrium tendency around the first magic angle ($\alpha\sim0.6$). Beyond that, and up to very strong couplings $\alpha\sim5$ (very low twist angles $\theta\sim0.1^{\circ}$), we observe, as noted in the main text, a complicated non-trivial evolution of the bandgap as a function of $\alpha$ and $\lambda$. This behavior is pronounced at nonzero $\lambda$. We particularly observe that the gap, at both Dirac points, can close at various points in the $\alpha,\lambda$ plane. In general, the gap at each Dirac point closes at different $\alpha$ and $\lambda$, but there are particular points at which both gaps become almost zero, reflecting an almost suppression of the perturbation effect in both layers (note, however, that the band structure profile would still be modified by the mass perturbation, even if the bandgaps become very small). In general, we found that the minima of the bandgap do not seem to correlate with the position of the magic angles in unperturbed TBG. 

The bandgap evolution is seen to depend also on the initial mass perturbation magnitude. For relatively small masses ($m<50\,\mathrm{meV})$, the profile of the bandgap evolution remains more or less similar. As the mass increases, we see that the gap oscillating behavior (namely, closing and reopening) only persist at large ratios $\lambda$. Eventually, at sufficiently large $m$ the bandgap evolution as $\alpha$ increase becomes more uniform. The nontrivial behavior of the bandgap can be attributed to the fact that as $\alpha$ increases the energy scale of the narrow bands becomes increasingly smaller, so the mass perturbation actually becomes stronger, leading to the complicated behavior observed.

The closure and reopening of the gap, as $\alpha$ increases beyond the first angle angle, implies charge transfers at the touching points and therefore changes in the topology of the bands. We have indeed observed that the valley Chern number of the isolated narrow bands changes according to the bandgap evolution, in a also seemingly nontrivial way. This striking feature could imply that topological changes --with Chern numbers beyond those conventionally expected for the system-- could be influenced (or externally induced) by asymmetrical perturbations in the coupled layers. 

\begin{figure}[t]
    \includegraphics[width=1\linewidth]{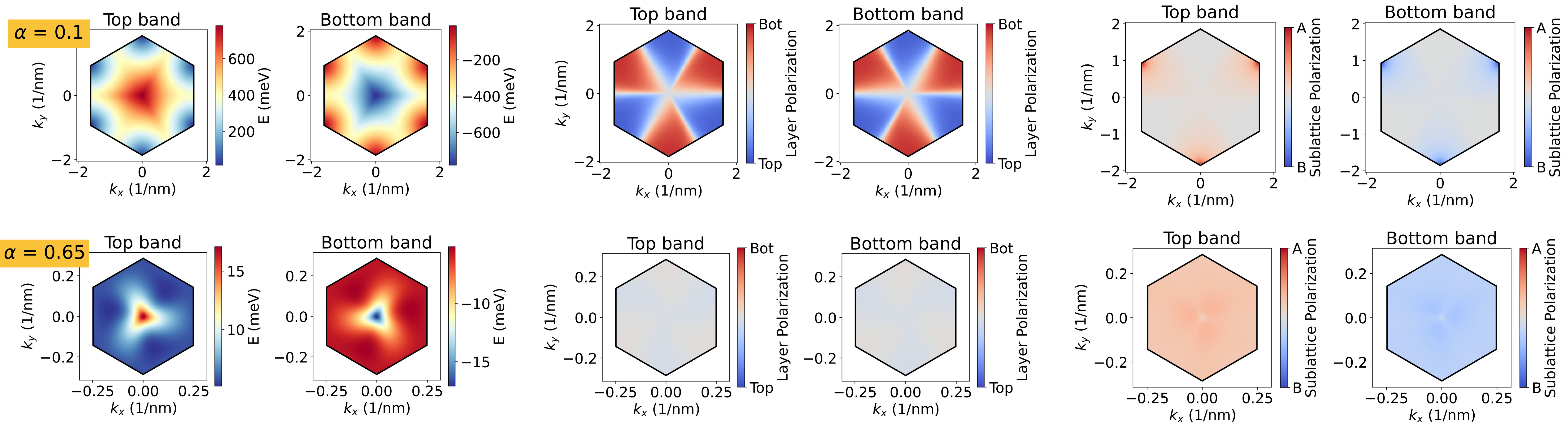}
    \caption{Energy and polarization density maps of the top and bottom central moiré bands, for the case of a mass perturbation $P_{b}=\sigma_{z}m$, $P_{t}=0$, and moiré coupling strengths $\alpha=0.1$ and $\alpha=0.65$ (equilibrium point). All cases correspond to $\lambda=0.8$. For each $\alpha$ it is shown the energy (left), layer polarization (middle) and sublattice polarization (right) of the top and bottom bands. 
 }\label{fig:mass_density}
\end{figure}

Figure \ref{fig:mass_density} shows energy and polarization density maps of the top and bottom central moiré bands, for the case of a mass perturbation $P_{b}=\sigma_{z}m$, $P_{t}=0$ (same as Fig. 2(a) in the main text). In the regime of low moiré coupling (e.g. $\alpha=0.1$), the two layers are weakly hybridized so the layer polarization is effectively top or bottom when close to the respective Dirac points. At the same time, since the gap is only pronounced in the perturbed bottom layer, the states close the bottom Dirac point are sublattice polarized. Upon reaching the moiré equilibrium at $\alpha=0.65$, the layer polarization tends to zero due to the strong layer hybridization. However, due to the bandgap equilibrium in both Dirac points, the top and bottom bands become sublattice polarized over the \emph{whole} moiré Brillouin zone. 

\begin{figure}[t]
    \includegraphics[width=1\linewidth]{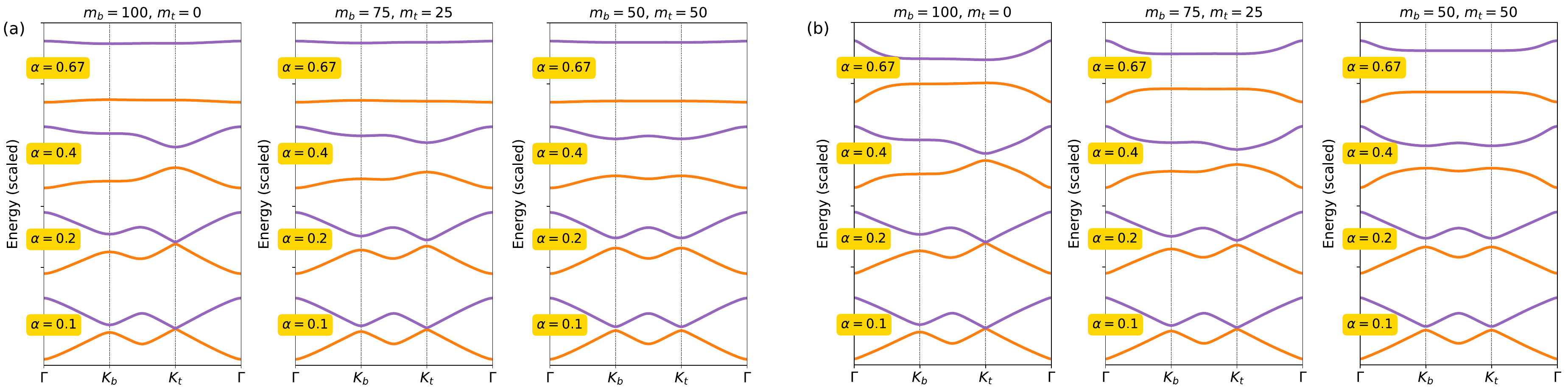}
    \caption{Evolution of the top and bottom narrow bands with (a) $\lambda=0$ (chiral limit) and (b) $\lambda=0.8$, for different coupling strengths $\alpha$, and total mass perturbation $m=100\,\mathrm{meV}$. For each $\lambda$, the three panels showcase different ways in which the perturbation is distributed among the layers, ranging from perturbation only in the bottom layer ($K_b$ point), to equally distributed perturbation in both layers. In either case, a moiré-driven equilibrium of equal bandgap at $K_b$ and $K_t$ is reached at $\alpha\approx0.67$.}\label{fig:BS_diff}
\end{figure}

Figure \ref{fig:BS_diff} shows the band structure of the top and bottom narrow bands at different moiré couplings, for (a) $\lambda=0$ and (b) $\lambda=0.8$, and three different scenarios: (i) perturbation only in the bottom layer, (ii) perturbation weighted $3/4$ and $1/4$ in the bottom and top layer, and (iii) equal perturbation in both layers. As seen, the bandgap at $K_b$ and $K_t$ tend to practically the same equilibrium regardless of how the initial mass perturbation is distributed in each layer. This behavior is consistent with the first-order perturbation result given by Eq. (12) in the main text, which indicates that the equilibrium tendency takes place independently of the perturbed masses $m_b$ and $m_t$. Note that when $m_{b}=m_{t}$ the band gap at both points is always equal, regardless of the moiré coupling strength $\alpha$, again in agreement with Eq. (12) in the main text. Figure \ref{fig:BS_diff} reflects that the moiré-driven equilibrium is effectively \emph{independent} of the initial perturbation in each layer.

\begin{figure}[t]
    \includegraphics[width=0.8\linewidth]{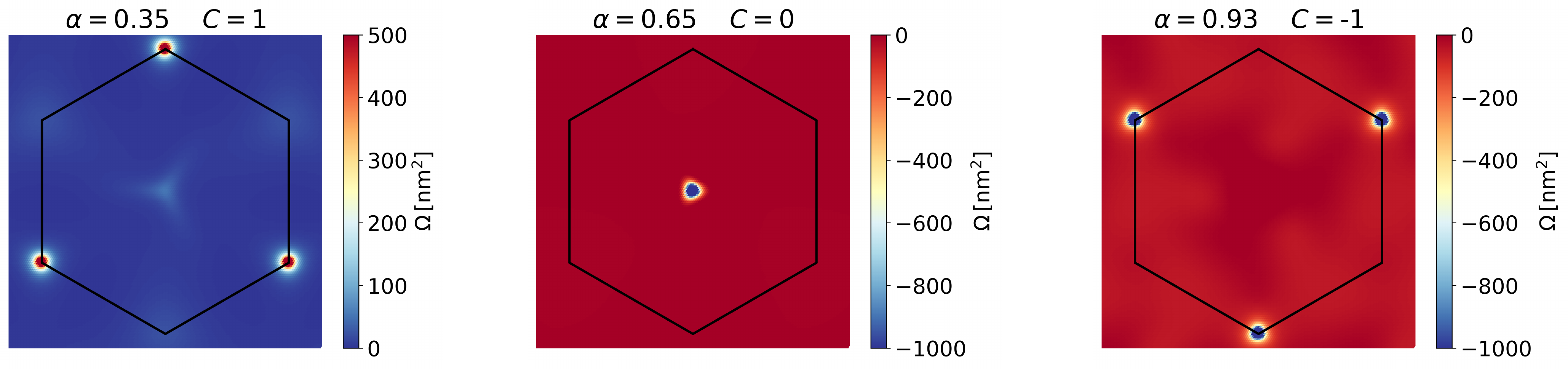}
    \caption{Berry curvature of the top narrow band, for a mass perturbation $m=50\,\mathrm{meV}$ on the bottom layer and three different moiré couplings $\alpha=0.36,0.65,0.93$. All other parametersare the same as in Figure 2 of the main text. In each coupling case, the scale of the Berry curvature is fixed for better visualization. The corresponding valley Chern numbers are $C=1,0,-1$, respectively. }\label{fig:berry_curv}
\end{figure}

As noted in the main text, the moiré-driven equilibrium under a mass perturbation also influences the Berry curvature \cite{xiao2010berry}. In general, the Berry curvature peaks near band degeneracies, such as Dirac points or points where the gap between two bands is smaller. This behavior is reflected in Figure \ref{fig:berry_curv}, which shows the Berry curvature of the top narrow band over the mBZ, for a mass perturbation $m=50\,\mathrm{meV}$ on the bottom layer (same configuration and parameters as Figure 2 in the main text). When coupling is small (e.g., $\alpha=0.35$), the Berry curvature is highly localized around the unperturbed Dirac point $K_{t}$, where the bandgap between the narrow bands is small. At the moiré equilibrium point $\alpha=0.65$, the gaps at both Dirac points become equal and so is the Berry curvature there. In this case, the Berry curvature rather becomes now concentrated at the $\Gamma$ point at which the gap between the top and remote band is smallest. Beyond the equilibrium point, the Berry curvature continues to redistribute. For instance, at $\alpha=0.93$ the gap at the unperturbed Dirac point now becomes very small, so the Berry curvature peaks strongly concentrated there. 

As noted before, these redistribution of the Berry curvature further implies changes in the (valley) Chern number (namely, whenever gaps between narrow, or between narrow and remote, close and reopen). The three cases illustrated in Figure \ref{fig:berry_curv} indeed reflect a transition from topological to trivial at the moiré equilibrium point, and again to topological beyond it. 

\subsection{Scalar perturbation}

\begin{figure*}[t]
    \includegraphics[width=1\linewidth]{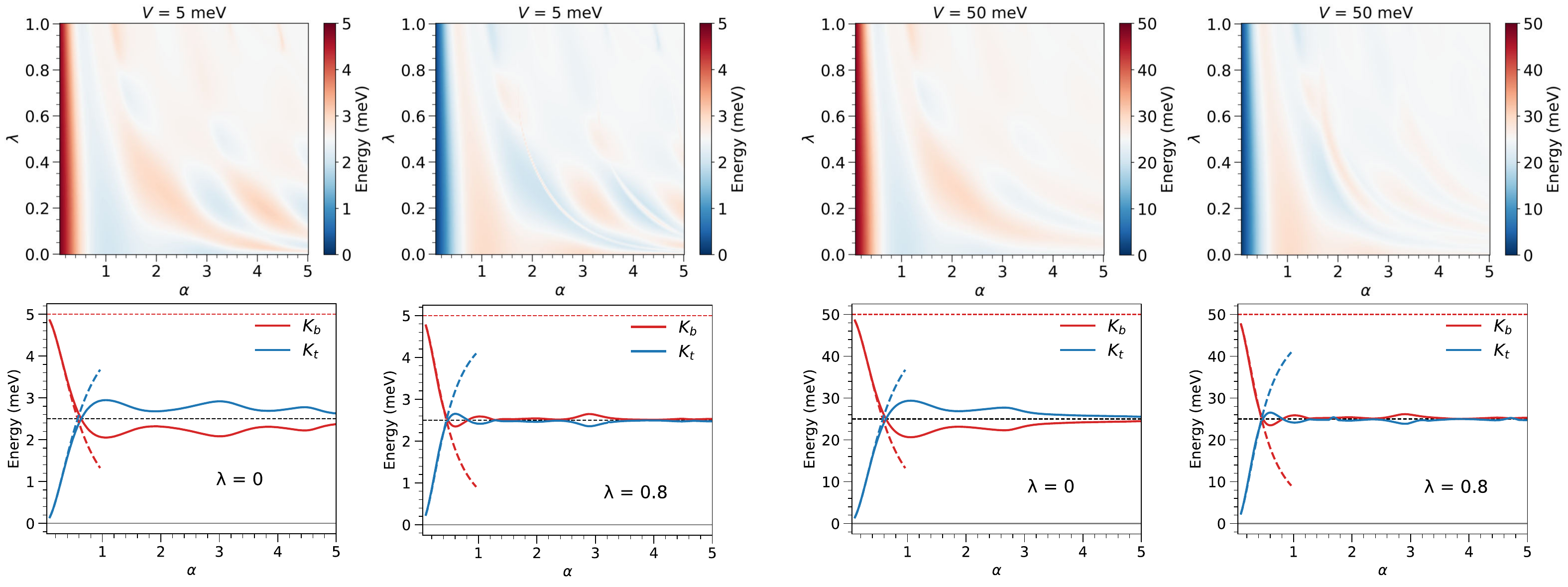}
    \caption{Evolution of the energy shift at the Dirac points, for a scalar perturbation $P_{b}=\sigma_{0}V,P_{t}=0$, with different magnitudes of $V$. For each $V$, the top two panels show density plots of the energy shift at the perturbed bottom layer (left, $K_b$ point) and unperturbed top layer (right, $K_t$ point), as a function of the moiré coupling strength $\alpha$ and ratio $\lambda$. Bottom panels show the energy shift at $K_b$ and $K_t$, as a function of $\alpha$ for $\lambda=0$ (left) and $\lambda=0.8$ (right). The horizontal red dashed-line indicates the decoupled energy shift $V$, while the black dashed-line indicates the equilibrium point $V/2$. Dashed curves correspond to the first-order perturbation result given by Eq. (8) in the main text.}\label{fig:scalar_SM}
\end{figure*}

Figure \ref{fig:scalar_SM} shows numerical results for a scalar perturbation $P_{b}=\sigma_{0}V$ in the bottom layer ($K_b$ point) with different values of $V$, and no perturbation in the top layer ($P_{t}=0$, $K_t$ point). As in Figure \ref{fig:bandgap_SM}, the top panels show a density map of the energy shift at the perturbed ($K_b$ point) and unperturbed ($K_t$ point) layers, while the bottom panel show the energy shift for $\lambda=0$ (chiral limit) and $\lambda=0.8$.

All cases reflect, once again, a moiré-driven equilibrium as $\alpha$ increases, independently of the perturbation strength. In line with the first-order result given by Eq. (14) in the main text, the equilibrium energy shift is \emph{always} half the initial shift ($V_{\mathrm{eq}}^{\star}=V/2$), regardless of the ratio $\lambda$. Beyond the first magic angle, the shift oscillates in a nontrivial way around the equilibrium value, similar to the bandgap behavior observed in Figure \ref{fig:bandgap_SM}. However, the oscillating behavior for the scalar perturbations is less pronounced than for the mass perturbation. For the chiral limit $\lambda=0$, the energy shift beyond the first magic oscillates but \textit{never} reaches again the equilibrium point, similarly to how the bandgap in Figure \ref{fig:bandgap_SM} oscillates when $\lambda=0$. When $\lambda\neq0$, however, the energy shift oscillates and touches the equilibrium point repeatedly as $\alpha$ increases, again following a similar pattern as the bandgap behavior for $\lambda\neq0$. 

\subsection{Gauge perturbation}\label{sec:app_gauge}

As noted in the main text, a notable manifestation of moiré equilibrium is the collapse of Dirac points under a gauge perturbation. Figure \ref{fig:DP_collapse} shows such collapse for six different gauge perturbation directions. As noted in the main text, the Dirac points collapse along the gauge perturbation direction only if such direction connects opposite high-symmetry points within the mBZ (three top panels in Figure \ref{fig:DP_collapse}). These three high-symmetry directions are determined by three transfer vectors in Eqs. \eqref{eq:q_transer}. For other perturbation directions, not aligned with the transfer vectors, the collision path bents in order to achieve the collision. 

For a single layer gauge perturbation, as considered in Figure \ref{fig:DP_collapse}, the first-order perturbation result predicts that the shift of the Dirac points (and consequently the collision path) is \emph{along} the direction of the gauge perturbation; see Eq. \eqref{eq:A_shift}. This result does not depend on the initial single-layer perturbation direction because it only holds within the isotropic Dirac cone approximation. The first-order perturbation captures the correct (full continuum model) collision path if the Dirac points can effectively collide by following the gauge direction, which can only happens along the high-symmetry directions determined by the transfer vectors. 

\begin{figure*}[t]
    \includegraphics[width=1\linewidth]{FigureS6.pdf}
    \caption{Collapse of Dirac points for a gauge perturbation $P_{b}=\boldsymbol{\sigma}\cdot\mathbf{A}_{b}$ acting on the bottom layer, with magnitude $\left|\mathbf{A}_{b}\right|=0.03\;\mathrm{nm^{-1}}$ and different directions. In each case, the small left panel shows the mBZ with the perturbation direction $\mathbf{A}_{b}$ acting on the bottom Dirac point (red dots), while the larger right panel shows the collision path of the top (blue) and bottom (red) Dirac points, as the moiré coupling increases from $\alpha=0.3$ to $\alpha=0.7$, in steps of $0.01$. The top three cases show the situations in which the perturbation direction, and the collision path, is along the high symmetry directions connecting opposite Dirac points. In the bottom three cases, the perturbation direction does not connect opposite Dirac points, and the collision path shift with respect to it. }\label{fig:DP_collapse}
\end{figure*}

\subsection{Strain perturbation}\label{sec:app_strain}

\begin{figure*}[t]
    \includegraphics[width=1\linewidth]{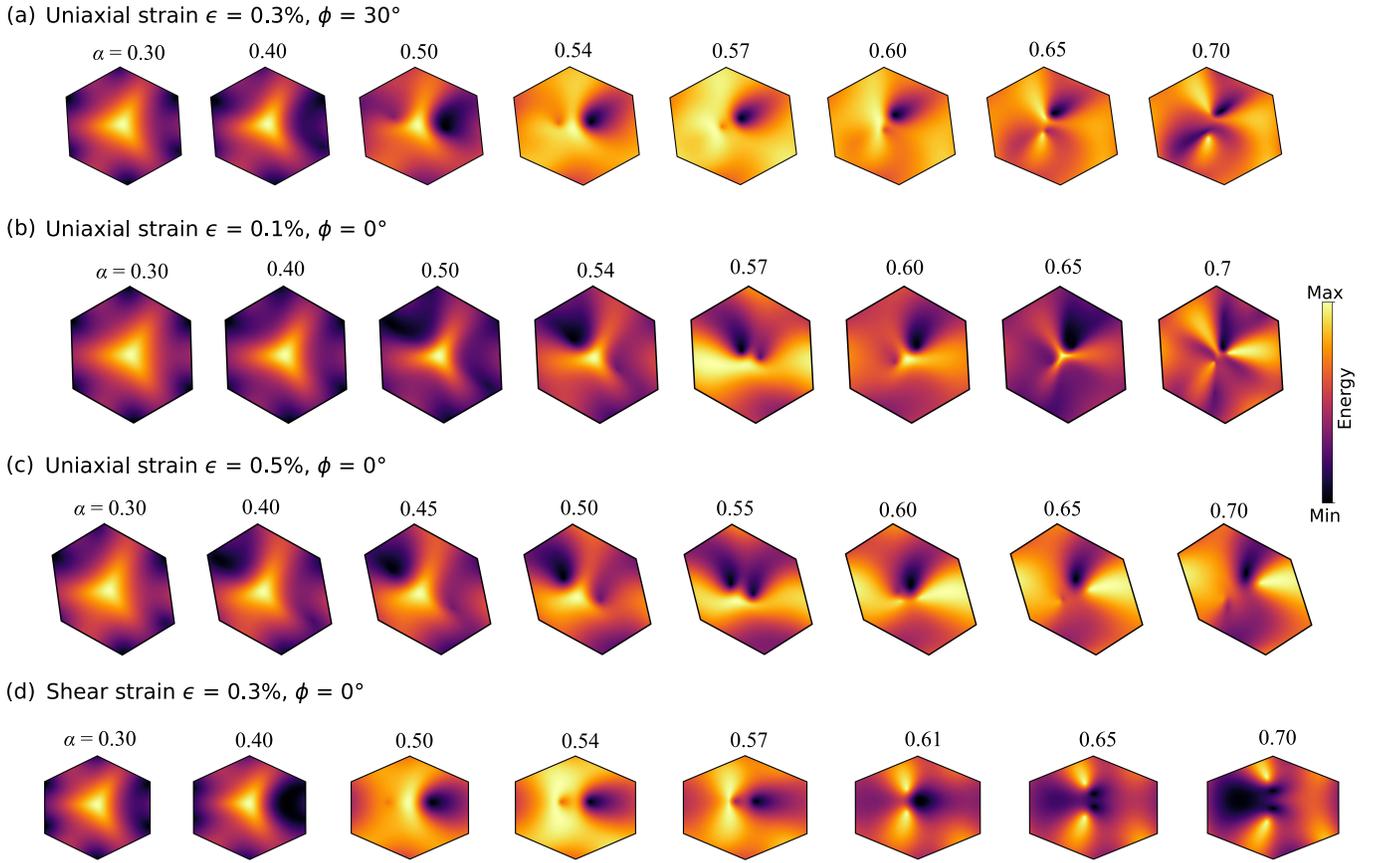}
    \caption{Evolution of the Dirac points for different moiré couplings $\alpha$ and strain configurations. Each case shows a density plot of the top narrow band within the moiré Brillouin zone. The moiré Brillouin zone is deformed due to the strain effect.}\label{fig:DP_strain_extended}
\end{figure*}

Figure \ref{fig:DP_strain_extended} shows the evolution of the Dirac points in the top narrow band, for different moiré couplings $\alpha$ (case $\lambda=0.8$) and strain configurations \cite{escudero2025geometrical}. All the results correspond to a heterostrain configuration in which only the top layer is strained \cite{huder2018electronic, mesple2021heterostrain}.

As in Figure 3 of the main text, in all cases shown we see an equilibrium tendency in the position of the Dirac points. Although with strain the moiré potential is also perturbed (Sec. \ref{app:strain}), the overall behavior remains the same: As the moiré coupling increases, the Dirac points shift their position towards a collapse. In most cases, an actual collapse does not technically occur because the perturbation difference $\mathbf{A}_{t}-\mathbf{A}_{b}$ is not aligned with the difference between the position of the (twisted and strained) Dirac points (see main text). The collapse possibility is also influenced by the perturbation of the moiré potential. What does persist in all cases is a moiré-driven equilibrium. 

\end{document}